\documentstyle[prl,aps]{revtex}
\begin{document}
\twocolumn[\hsize\textwidth\columnwidth\hsize\csname@twocolumnfalse\endcsname
\author{A.E.Koshelev$^{1,2}$, P.Le Doussal$^3$, and V.M.Vinokur$^1$}
\title{Columnar defects and vortex fluctuations in layered superconductors}
\date{\today}
\address{$^1$ Materials Science Division, Argonne National
Laboratory,Argonne, IL 60439\\
$^2$ Institute of Solid State Physics, Chernogolovka,
Moscow,142432, Russia\\
$^3$ CNRS- Laboratoire de Physique Th\'eorique de l'Ecole\\
Normale Sup\'erieure, 24 rue Lhomond, F-75231 Paris}
\date{\today}
\maketitle
\begin{abstract}
We investigate fluctuations of Josephson-coupled pancake vortices in
layered superconductors in the presence of columnar defects.  We study the
thermodynamics of a single pancake stack pinned by columnar defects and
obtain the temperature dependence of localization length, pinning energy
and critical current.  We study the creep regime and compute the crossover
current between line-like creep and pancake-like creep motion.  We find
that columnar defects effectively increase interlayer Josephson coupling by
suppressing thermal fluctuations of pancakes.  This leads to an upward
shift in the decoupling line most pronounced around the matching field.
\end{abstract}
\vspace{0.25in}
\twocolumn
\vskip.5pc]
\narrowtext
The question of the effective dimensionality of the vortex structure
in highly anisotropic high-$T_c$ superconductors is the subject of
intense experimental and theoretical interest.  It was predicted that
thermal fluctuations of vortex positions produce strong relative phase
fluctuations in adjacent layers and that the phase coherence along the
$c$-axis is destroyed above the decoupling line (DL) in the $H-T$
plane \cite{GlKosh}.  Above this line an additional transverse field
is not screened by Josephson currents and completely penetrates the
sample.  Several experiments were interpreted \cite{decoupl} as a
manifestation of decoupling effects.

In this Letter we focus on the influence of columnar defects produced by
heavy ion irradiation on the vortex decoupling and the resulting effective
dimensionality of the vortex structure. The interest in this question is
motivated not only by the technological importance of highly anisotropic
compounds such as $Bi_2Sr_2CaCu_2O_x$ (BSCCO), but also by
the expected rich variety of nontrivial dynamic and static properties
arising from the interplay between the continuous linear geometry of heavy
ion tracks and the discrete pancake-like character of the vortex structure.

An important question is whether columnar defects can be especially
efficient to pin 2D pancakes which are nearly independent in each
layers.  Early experiments \cite{2dexp} of irradiation of BSCCO failed
to show unidirectional pinning along the columns thus suggesting that
the linear nature of columnar defects may not be efficient for highly
anisotropic materials.  However subsequent experiments revealed
\cite{ColDefExp,lineexp,Kees} that the {\it relative} enhancement of
critical current (by several orders of magnitude) and the upward shift
of the technologically useful irreversibility region, was even {\it
greater} for highly anisotropic material such as BSCCO than for the
less anisotropic YBCO.

We demonstrate that the observed enhanced pinning of pancakes can be
explained in terms of the {\it recoupling} of the layers by the
columnar defects.  We show that columnar defects effectively increase
the interlayer Josephson coupling by suppressing thermal fluctuations
of pancakes positions.  This leads to an upward shift in the
decoupling line, the maximal shift being expected around the matching
field.  We also investigate in detail pinning regimes of a single
pancake stack on a single column and on the system of columns.

The theory of vortex pinning by columnar defects in anisotropic 3D
superconductors was developed in \cite{NelsVin}.  Mapping vortex lines onto
the imaginary-time quantum mechanics of 2D bosons subject to static
disorder leads to the prediction of a novel low temperature ''Bose glass''
phase with vortices localized near columnar pins, an infinite tilt modulus
and zero linear resistivity.  The Bose glass transition line $B_{BG}(T)$ is
shifted to higher temperatures as compared to the melting line $B_m(T)$ of
the clean lattice.  The existence of the Bose glass phase is supported by
several experiments \cite{BoseGlExp,Kees}.  A new feature of highly
anisotropic superconductors is the discrete nature of vortices.  The
physics of the system is governed by three characteristic lengths: the
vortex spacing $a=(\Phi_0/B)^{1/2}$, the distance between defects $d$ and
the Josephson length $\gamma s$.  Here $B$ is the magnetic induction,
$\Phi_0$ is the flux quantum, $s$ the interlayer spacing and $\gamma$ is
the anisotropy parameter.  The Josephson length defines the crossover field
$B_{cr}=\Phi_0/(\gamma s)^2$ above which the interplane interaction energy
of pancakes exceeds the interlayer coupling.  The discreteness changes the
depinning temperature, localization length, binding energy, and critical
current.  We calculate these parameters using a discrete-time
generalization of the mapping onto quantum mechanics.  We show that at low
temperature and moderate fields the system crosses over from pancake to
line-like behaviour.  We study this crossover at short scales in
thermodynamic quantities and in transport properties as a function of the
current.  At elevated temperatures and fields the layered nature of the
superconductor is essential and leads to a new phase, the ``decoupled Bose
glass'', where the phase coherence along the $c$-axis is lost but pancakes
are aligned by columns and therefore the system exhibits infinite tilt
stiffness.

The statistical mechanics of a single stack of pancakes near a columnar
defect is determined by the partition function:
\begin{eqnarray}
Z_N({\bf r}_0,{\bf r}_N)&=&
\int\limits_{r_n>b_0}\prod_{0<n<N}{\frac{\epsilon_1d{\bf r}_n}{2\pi sT}}\\
\nonumber&&\exp \left( -{\frac 1T}\sum_{n=1}^N\left[
{\frac{\epsilon _1}{2s}}({\bf r}_n-{\bf r}_{n-1})^2+V({\bf r}_n)\right]
\right)
\end{eqnarray}
depending on the positions of the start and end points ${\bf r}_0,{\bf
r}_N$\cite{don}.  The Josephson coupling binds the pancakes in the
neighboring layers by a spring-like interaction with the elastic
constant $\epsilon _1=(\epsilon _0/\gamma ^2)\ln ({\gamma s/}r_{cut})$
where $\epsilon _1$ is the elastic tension in the continuous limit,
$\epsilon _0=\Phi _0^2/(4\pi \lambda )^2$, $\lambda $ is the London
penetration depth for the currents flowing along $ab$ planes and
$r_{cut}$ being the cut off radius of order of $b_0$ if
$r_n,r_{n-1}\lesssim 2b_0$ and of order of $|r_n-r_{n-1}|$ otherwise.
We assume $\gamma <\lambda /s$, and therefore neglect the magnetic
coupling as compared to the Josephson one.  We model columnar defects
as cylinders of normal material of the radius $b_0$.  The pancakes
interaction with the columnar defect is given by an attractive pinning
potential $V(r)$, which at $r-b_0\gg \xi $ can be found from the
London equations \cite{VVShmidt},
\begin{equation}
V(r)=s\epsilon _0\ln \left(1-b_0^2/r^2\right)  \label{ColPot}
\end{equation}
The maximum pinning potential per unit length is $U_0\approx \epsilon
_0\left( \ln (b_0/\xi )+0.38\right)$.

Using the standard transfer matrix method one finds that in the large-$N$
limit $Z_N({\bf r},{\bf r}^{\prime })$ takes the form:
\begin{equation}
Z_N({\bf r},{\bf r}^{\prime })=\exp (-E_0sN/T-V({\bf r}^{\prime })/T)\phi
_0({\bf r})\phi _0({\bf r}^{\prime })
\end{equation}
where $E_0$ (the binding energy per unit length) and $\phi _0({\bf r})$ are
the ground-state energy and the ground state wave function of the discrete
analogue of a 2D ''Shr\"odinger equation'':

\begin{eqnarray}
\exp({-E_0s/T}) \phi _0({\bf r})&=&\\
\nonumber
\int {\frac{{\epsilon _1}d {\bf r}^{\prime }}{{2\pi sT}}} &\exp&
\left( -{\frac{{\epsilon _1}({\bf r}-{\bf r}^{\prime
})^2}{{2sT}}}-{\frac{{V({\bf r}^{\prime })}}T}\right) \phi _0( {\bf
r}^{\prime }).
\label{EqPhi}
\end{eqnarray}
Taking the limit $s\to 0$ recovers the usual quantum mechanical
representation of vortex line. The probability of finding a pancake at
point ${\bf r}$ near columnar defect is:
$P({\bf r})=\exp (-V({\bf r})/T)\phi _0^2({\bf r}) $

Localization effects are governed by the effective potential $U_{eff}({\bf
r})=-(T/s)\left( \exp [-V({\bf r})/T]-1\right)$ which is characterized by
the large dip in the close vicinity of the column and has a $1/r^2$ tail far
from the defect. The small distance dip dominates at low temperatures while
the large distance tail becomes important at high temperatures (the exact
criterion will be specified below).
At low temperatures~the~potential is effectively short-ranged and can be
approximated by a delta-function $U_{eff}({\bf r})\approx -W\delta ({\bf
r})$ with $W=(2\pi T/s)\int_{r>b_0}rdr(\exp [-V(r)/T]-1)$.
The main contribution to $W$ at low temperatures comes from short distances
and can be estimated as $\left( 2\pi Tb_0r_0/s\right) \exp (sU_0/T)$, with
the length $r_0$ being dependent on the detailed mechanism of the pancake
nucleation at the column boundary. Approximating the pancake potential at
distances $0<r-b_0\lesssim \xi $ by $V(r)\approx sU_0+(s\epsilon _0/2)\ln
(1+(r-b_0)^2/\xi ^2)$ we obtain $r_0\approx \xi (T/s\epsilon _0)^{1/2}$.
Within the delta-function approximation Eq. (\ref{EqPhi}) can be solved
exactly \cite{us} giving the ground state energy as:
\begin{equation}
E_0=(T/s)\ln \left\{ 1-\exp \left( -{2\pi }T^2/(\epsilon _1W)
\right) \right\} .  \label{GrStEn}
\end{equation}
The distribution function is then $P({\bf r})=\phi _0^2( {\bf
r})+\left( sW/T\right) \phi _0^2(0)\delta ({\bf r})$, i.e., it
consists of the delta-peak centered at the defect having the weight
$1-\exp (sE_0/T)$ and the smooth part with the weight $\exp (sE_0/T)$.
A similar structure for $P({\bf r})$ has been observed in numerical
simulations \cite{don} .  We define the depinning temperature $T^{*}$
by the condition that at $ T>T^{*}$ the probability to find a pancake
outside the column becomes higher than the probability to find it
inside the defect.  Matching probabilities of finding pancake inside-
and outside the defect gives rise to the self-consistent equation
\begin{equation}
T^{*}=sU_0/\ln \left( {\frac{{sT^{*}}}{{b_0r_0\epsilon _1}}}\right) ,
\label{DepTemp}
\end{equation}
where the quantities $U_0$ and $\epsilon _1$ are taken at $T=T^{*}$. Eq.~(
\ref{DepTemp}) is valid provided $\xi _l(T^{*})\gg b_0$, this is equivalent
to the condition $\gamma s\gg b_0$ and defines a range of parameters
corresponding to the ``pancake'' regime of depinning as opposed to the
``elastic string'' regime described in Ref. \cite{NelsVin}. Above $T^{*}$
the localization parameters obey the ''exponent-in-exponent'' temperature
dependencies, in particular, for the localization length $\xi _l$, $\xi
_l^2=\int d^2{\bf r\;r}^2\phi _0^2({\bf r})$, one has
\begin{equation}
\xi _l^2\approx {\frac{{sT}}{3{\epsilon _1}}\exp }\left[ \frac{sT\exp
(-sU_0/T)}{\epsilon _1b_0r_0}\right] .  \label{LocLength1}
\end{equation}
Evaluating the parameter $W$ with the potential from (\ref{ColPot}) one
finds that the $1/r^2$ tail gives rise to a logarithmic divergence at large
distances which has to be cut off at $\xi _l$. Finally one arrives at $
W\approx \frac{2\pi }sTb_0r_0\exp \left( {\frac{{sU_0}}T}\right) +{\pi
b_0^2{
\epsilon _0}}\ln {\frac{{\xi _l^2T}}{{\ s\epsilon _0b_0^2}}}$ , where the
first term is due to the short distances contribution. Substituting the
expression for $\xi _l$ from Eq. (\ref{LocLength1}) into the last equation
we can see that the tail region dominates above the typical temperature $
T^{**}=2sU_0/\ln \left( {s^2U_0/b_0^2\epsilon _1}\right) >T^{*}$.
Numerical estimate for BSCCO with
$\lambda_0=1800 \AA$, $\gamma=200$ $b_0=100\AA $
gives $T^*\approx 69$ K, $T^{**}\approx 73$ K.

At temperatures significantly larger than $T^{**}$ discreteness is not
important any more, Eq.( \ref{EqPhi}) reduces to the Shr\"odinger equation,
and the localization parameters coincides with those for the elastic
string
\cite{us,Blatter}
\begin{equation}
\xi _l^2=\gamma sb_0\exp \left(
\frac T{T_{\rm str}}\right) ;\;E_0={\frac{T^2}{T_{\rm str}b_0}}\exp \left(
-\frac
T{T_{\rm str}}\right) .  \label{HighTE0}
\end{equation}
with $T_{\rm str}=\sqrt{2\epsilon _1\epsilon _0}b_0/\pi $. The single-column
regim  breaks down at temperature $T_{dl}$ where the localization length $
\xi _l$ matches the spacing between the columns, $\xi (T_{dl})=d$.

Next we discuss transport of the pancake vortices in the presence of
columnar defects at low temperatures and fields ($B<B_\phi =\Phi_0/d^2
<B_{cr}$) below critical current.  The pancake
motion in this region is controlled by the liberation of a single
stack of pancakes from the localized state near a columnar defect.
The energy cost to liberate a segment of $N$ pancakes is:
$$
\Delta E_N =\sum_{n=1}^N\left( {\frac{{\epsilon _1}}{2{s}}}
(r_n-r_{n+1})^2+sU_0+V(r_n)\right) -\frac{\Phi _0}cjsr_n
$$
The barrier $V$ is determined by the lowest energy saddle point
configuration determined by the stationarity condition $\delta \Delta
E_N/\delta r_n=0$.  One can distinguish the string-like and single
pancake regimes separated by the ``decoupling'' current $j_{\rm
dec}=c\sqrt{\epsilon _1U_0}/(\Phi _0s)$.  At $ j<j_{\rm dec}$ the
critical nucleus consists of $N>1$ pancakes .  The continuum limit
corresponds to $N\gg 1$, i.e $j\ll j_{\rm dec}$ and one finds $
N\simeq 2\sqrt{2}j_{\rm dec}/j>>1$.  The resulting string-like creep
is characterized by a half loop width $w\approx cU_0/\Phi _0j$ and the
barrier $V_{\rm str}(j)\approx 4\sqrt{2}sU_0 j_{\rm dec}/(3j)$ and
holds as long as $w<a$ ,$d$.  Because of discreteness the number of
the pancakes in the nucleus decreases by one with increase of the
current at well defined transition currents $j=j_{N\rightarrow N-1}$,
$j_{N\rightarrow N-1}=j_{\rm dec}\sqrt{2\sqrt{2}/N(N+1)}$.
Thus for $j_{N+1\rightarrow N}<j<j_{N\rightarrow N-1}$, the barrier
$V(j)$ is determined by the lowest energy for a $N$ pancake
excitation: $ V(j)=E_N(j)=U_0\left( N-\frac{j^2}{24j_{\rm
dec}^2}N(N+1)(N+2)\right) $.  For $j\ll j{\rm dec}$ the last
expression reduces to $V_{\rm str}$.  Above $j_{\rm dec}$ pancakes
liberate from the column independently with activation energy
$\approx sU_0$.

We turn now to the collective localization of a pancake stack on the
forest of columnar defects for $B<B_{cr}<B_\Phi $.  There is a new
``tight binding'' regime for a localization of a single pancake stack
by many columnar defects, which does not exist for continuous lines.
Above the temperature $T_J\approx \epsilon _1d^2/s$ pancakes can
travel from column to column, but as long as
$T<T^{*}$ one can neglect the probability to find pancakes
outside the columns.  There are two sources for
weak localization on the forest of identical columns: fluctuation of column
density and potential fluctuations
due to superposition of individual long range column potentials.
Here we focus on the density fluctuations which dominates at
high temperatures, $T>s\epsilon_0b_0/d$ and consider
potential fluctuations elsewhere \cite{us}.  The
localization length $\xi _l\gg d$ can be estimated as follows.
When confining the pancake stack within the
region of size $\xi _l$ with enhanced number of columns $N=\bar
N+\delta N$, $\bar N\approx (\xi _l/d)^2$ and $\delta N\sim \sqrt{\bar
N}$, we gain the positional entropy due to the enhanced number
of available states as $-(T/s)(\delta N/{\bar N} )\approx -(T/s)(d/\xi
_l)$ .  The entropy loss due to suppression of
pancake thermal wandering by columnar defects is $ T^2/\epsilon _1\xi
_l^2$.  Optimizing the change in free energy $\delta
F=-(T/s)(d/\xi _l)+T^2/\epsilon _1\xi _l^2$ we find
the localization length and binding energy:
\begin{equation}
 \xi_l\approx Ts/\epsilon_1d;\ E_b\approx -\epsilon _1(d/s)^2.
\label{LocPar}
\end{equation}
It is important to note that in this regime the effectiveness
of localization decreases with increase of column concentration due
to the decrease of elemental jump distance.

The above results allow us to estimate the typical current,
$j_{\rm dec}\approx c E_b/(\Phi _0\xi _l)$ $\approx
c\epsilon_1^2 d^3/(\Phi_0s^3T)$ at which the stack of pancake delocalizes
from the weakly localized state.  Note however that at $T<T^{*}$ the
barriers controlling vortex hopping between columns of the are still
large (of order of $sU_0$), and thus for $j>j_{\rm dec}$ the resistivity
still remains much smaller than the flux flow resistivity $
\rho _{ff}$: $\rho \approx \rho _{ff}\exp (-{\frac{{sU_0}}T})$. The
resistivity approaches $\rho _{ff}$ at much larger current $j_c=cU_0/(\Phi
_0\xi )$. The current $j_{\rm dec}$ plays a role of ``decoupling current''
because at $j>j_{\rm dec}$ the interlayer coupling does not influence much
the motion of pancakes.

At elevated temperature pancakes spend more time outside the columns
and at $T=T^*$ there is a crossover from the ``tight binding'' regime
to the gaussian random potential localization.  In the latter regime
vortices are localized by the gaussian fluctuations of the effective
potential $\delta U_{eff}\approx -(T/s)\left[
\exp\left(sU_0/T\right)-1\right]b_0r_0/(\xi_l d)$. Adding the
confinement term $T^2/\epsilon _1\xi _l^2$ and optimizing
leads to:
\begin{equation}
\xi_l={\frac{sTd}{\epsilon_1b_0r_0}} (e^{\frac{sU_0}{T}}-1)^{-1};
E_b=-{\frac{\epsilon_1b_0^2r_0^2}{sd^2}}
(e^{\frac{sU_0}{T}}-1)^2
\label{aboveTstar}
\end{equation}

Eqs.~(\ref{LocPar},\ref{aboveTstar}) are valid at very small fields when
intervortex interactions can be neglected.  Interactions become
relevant for $B > B_{cp}\approx \Phi_0 E_b/(\epsilon_0\xi_l^2)$ and lead to a
collective pinning regime \cite{us}.  For the ``tight binding'' regime
the crossover to collective pinning is expected at $B_{cp}\approx
B_{cr} T_J^2 /T^2$.

We now discuss the suppression of thermal decoupling by columnar
defects.  The effects of columns on the phase coherence between the
layers is of a very different nature at high and low fields and have
to be discussed separately.  In the absence of columnar defects and at
small fields $B<B_{cr}=\Phi_0/(\gamma s)^2$ the decoupling occurs in
the {\it liquid} phase when the mean square fluctuation, $u_{\rm neib}^2=
\left\langle \left( r_{n+1}-r_n\right) ^2\right\rangle $, of the
separations between the adjacent pancakes belonging to the same vortex
line becomes of order $a^2$ \cite{GlKosh}.  For pin-free system
$u_{\rm neib}^2=2sT/\epsilon _1$ and $T_{\rm dec0}(B)\sim \epsilon
_1\Phi _0/2sB\sim T_m(B_{cr}/B)^{1/2}>T_m$
where $T_m$ is the melting temperature of the pure system. Columnar defects
suppress
the tilt deformations of vortex lines and reduce their influence on
the interlayer Josephson coupling.  Since the vortex-vortex
interactions have little effect on $u_{\rm neib}^2$ the above result
for a single stack can be used.  Correction to $u_{\rm neib}^2$ caused
by columns is connected to binding energy by relation $\Delta u_{\rm
neib}^2 =2s^2(\partial E_b/\partial \epsilon _1)$.  For temperature
range $T<T^*$ decoupling takes place within the ``tight binding''
regime.  Using (\ref{LocPar}) we obtain $u_{\rm neib}^2 \approx
2sT/\epsilon _1-\hbox{Const }d^2$ and the Lindemann-like criterion
$u_{\rm neib}^2 \approx a^2$ gives
\begin{equation}
T_{\rm dec}\approx T_{\rm dec0}+\hbox{Const }\epsilon _1d^2/s, \hbox{ at
} T_{\rm dec}< T^*
\label{DecSmB}
\end{equation}
Note that in this regime the decoupling temperature {\it decreases}
with the increase of the concentration of the columns due to decrease
of the elemental jump distance.  At $T_{\rm dec}>T^*$ one finds, similarly:
\begin{equation}
T_{\rm dec}\approx T_{\rm dec0}
+{\frac{sb_0^2r_0^2}{d^2}}\left[\exp{\frac{sU_0}{T_{\rm
dec0}}}-1\right]^2.
\end{equation}
As temperature approaches $T_{c}$, the renormalized decoupling line
smoothly merges the unrenormalized one.

At large fields $B\gg B_\Phi ,B_{cr}$ one has to take into account that (i)
only a small fraction of vortices $\approx B/B_\Phi$ is trapped by
columns, and
(ii) the interaction between the vortex lattices in different layers is
weaker than intralayer interactions. To estimate fluctuations of the phase
difference between layers we note that the phase disturbance is $\delta
\phi _n({\bf q})=2\pi n_vu_{nt}/q_{\perp }$, where $u_{nt}$ is
the transverse component of the lattice displacement in the $n$-th
layer at the wave vector ${\bf q}$. The resulting average phase
difference between the layers $\left\langle (\Delta \phi )^2\right\rangle
\equiv \langle (\delta \phi _n({\bf r})-\delta \phi _{n+1}({\bf r}
))^2\rangle $ is
\begin{equation}
\left\langle (\Delta \phi )^2\right\rangle =(2\pi n_v)^2\int_{q_{\perp
}>\pi /\gamma s}\frac{d{{\bf q}_{\perp }}}{{(2\pi )^2}}\frac{\langle
|u_{nt}(q_{\perp })|^2\rangle }{{\ }q_{\perp }^2}{.} \label{DPhi}
\end{equation}
The amplitude of the transverse vortex displacement can be
estimated as $\langle |u_{nt}(q_{\perp })|^2\rangle \approx T/({
sC_{66}(q_{\perp }^2+}\left( {\pi /d}\right) ^2))$, where $C_{66}=B\epsilon
_0/4\Phi _0$ is the shear modulus.  Performing
integration we find $
\left\langle (\Delta \phi )^2\right\rangle \approx TB/(s\epsilon
_0B_\Phi ) {\ln }(\gamma s/d)$.  Thus, the typical temperature $T_{\rm
dec}$, above which the Josephson coupling is considerably suppressed
can be estimated as
\begin{equation}
T_{\rm dec}\approx \frac{B_\Phi s\epsilon _0}{B{\ln }\left( \gamma
s/d\right)}.
\label{Tdc}
\end{equation}
The ratio of $T_{\rm dec}$ to the decoupling temperature for the ideal
crystal $T_{\rm dec0}$ \cite{GlKosh}, $T_{\rm dec0}=\left( \Phi _0/4\pi
B\right) ^{1/2}\epsilon _0/\gamma $, is
\begin{equation}
T_{\rm dec}/T_{\rm dec0}\approx B_\Phi {/}\sqrt{B_{cr}B}
\end{equation}
Thus,  at fields $B<B_\Phi^2/B_{cr}$ the columnar defects effectively
increase the coupled region.
For $B>B_\Phi^2/B_{cr}$ the curve $T_{\rm dec}$ smoothly merges with
the decoupling line of the pure system.
An important observation is that at large fields
$T_{\rm dec}(B)<T_m$  and approaches
$T_m$ at $B\approx B_\Phi$.
The vortex phase in the region $T_{\rm
dec}(B)<T<T_m$ has infinite tilt modulus and
therefore can be referred to as the ``decoupled Bose glass''. The largest
upward shift of the decoupling temperature is expected to occur
at $B\approx B_\Phi$ where
the thermal motion of pancakes is maximally suppressed
and pancakes do not travel from column to column.

To conclude, we have found three regimes of pancake localization on
one column: low temperature, ``pancake depinning'', and ``string
depinning'' regimes.  A new ``tight binding'' regime of localization of
a single pancake stack on many columns was found at high
concentration of columns.  We have studied the dynamic crossovers
from line to pancake creep, which can be observed in transport
experiments.  We have shown that columnar defects ``recouple''
Josephson layers by suppression of pancake positional fluctuations.
This effect can be checked in neutron diffraction and $\mu$SR
experiments.  We predict an upward shift of the decoupling
line by columnar defects.  A ``decoupled Bose glass'' phase is
predicted at high field.

VMV and AEK acknowledge support from Argonne National Laboratory
through the U.S.  Department of Energy, BES-Material Sciences, under
contract No.  W-31-109-ENG-38.  AEK and PLD acknowledge support
from National Science Foundation Office of the Science and Technology
Center under contract No.  DMR-91-20000. AEK and VMV acknowledge
ISI Foundation in Torino where part of this work was completed.

\end{document}